\documentclass[a4paper,onecolumn,11pt,unpublished]{quantumarticle}
\pdfoutput=1

\usepackage[utf8]{inputenc}
\usepackage[english]{babel}
\usepackage[T1]{fontenc}
\usepackage{amsmath}
\usepackage{hyperref}

\usepackage{tikz}
\usepackage{lipsum}

\begin{document}

\title{Frequency shifts heralding ground state squeezing and entanglement of two coupled harmonic oscillators}

\author{Safoura Mirkhalaf}
\affiliation{Centre for Quantum Optical Technologies, Centre of New Technologies, University of Warsaw, Banacha 2c, 02-097 Warszawa, Poland}
\orcid{0000-0002-4154-8465}

\author{Helmut Ritsch}
\orcid{0000-0001-7013-5208}
\author{Karol Gietka}
\orcid{0000-0001-7700-3208}
\email{karol.gietka@uibk.ac.at}

\affiliation{Institut f\"ur Theoretische Physik, Universit\"at Innsbruck, Technikerstra{\ss}e\,21a, A-6020 Innsbruck, Austria} 
\maketitle

\begin{abstract}
 It is often argued that two linearly coupled quantum harmonic oscillators, even when cooled to their ground state, display no inherently quantum features beyond quantized energy levels. Here, we challenge this view by showing that their classical observables encode genuinely quantum features. In particular, we demonstrate that the characteristic frequency shifts observed in coupled oscillators signal non-classical correlations and ground-state entanglement at zero temperature corresponding to two-mode squeezing between the uncoupled modes. From a complementary perspective, these two effects, frequency shifts and squeezing, represent the same underlying phenomenon but expressed in different mode bases. What appears as a spectral renormalization in one description manifests itself as entanglement in the other. Frequency shifts therefore constitute an entanglement witness accessible via standard spectroscopy. While the underlying squeezing is not directly measurable, it can be exploited to enhance the signal-to-noise ratio in precision frequency measurements of individual oscillators without requiring squeezed quantum noise. This uncovers a new route to quantum-enhanced sensing within systems traditionally regarded as classical, offering fresh insight into how signatures of quantumness persists across the quantum-to-classical boundary.
\end{abstract}

\section{Introduction}
Coupled harmonic oscillators are among the most fundamental and commonly used systems in physical modeling~\cite{PhysRev.175.286,Kim2005,PhysRevE.97.042203}. They serve as minimal but solvable models for a wide range of physical phenomena~\cite{Kim2007}, from vibrational modes in molecules~\cite{Wilson1955,Ikeda1999,Fillaux2005,Delor2017}, phonons in solids~\cite{Girvin2019} through biophysics~\cite{Halpin2014,Fuller2014,Romero2014} to cavity modes in quantum optics~\cite{Law1994,Walther2006,aspelmeyer2014,PhysRevA.110.063703} and circuit quantum electrodynamics~\cite{Blais2021}. Due to their analytic tractability and ubiquity~\cite{oriol2019quadratic}, they played a central role in developing both classical and quantum theories of motion, decoherence~\cite{Caldeira1983,Feynman2000,Weiss2011}, and measurement~\cite{Braginsky1992,Gardiner2004, Wiseman2009}. When two quantum harmonic oscillators are coupled linearly, the resulting system remains Gaussian and exactly solvable~\cite{Adesso2007,Weedbrook2012,Serafini2017,HO_2025_quant,PhysRevA.103.023707,Heib_2025}, leading to the widespread conjecture that this dynamics is, in essence, classical~\cite{Walls1994,Eisert2003,Scully1997}. Indeed, many aspects of the coupled dynamics—such as normal mode formation, energy exchange, and resonance shifts—can be derived entirely from the framework of classical physics~\cite{MarionThornton}. As a result, coupled oscillator systems are commonly viewed as belonging to the classical domain unless explicitly driven into non-Gaussian regimes or subjected to strong measurements that reveal nonlocal correlations. This perception has contributed to the prevailing notion that linearly coupled quantum oscillators are quantum in name only, lacking the genuinely non-classical features as entanglement or squeezing. Hence they are frequently used as benchmarks or pedagogical examples where quantum and classical predictions coincide~\cite{Feynman1965,Breuer2002} and the quantum nature of the system isirrelevant.

In this work, we challenge this prevailing view by showing that coupled quantum harmonic oscillators, even when appearing classical in their dynamical behavior, can harbor deeply quantum correlations. Specifically, we reveal that the frequency shifts arising from the coupling can serve as indirect signatures of two-mode squeezing and entanglement between the original uncoupled modes. These correlations, though not directly measurable in standard normal mode observables, leave a classical imprint on the system's response, allowing the frequency shifts to be used as entanglement witnesses. Furthermore, we show that the inherent latent squeezing can be harnessed to enhance the precision of frequency estimation for the individual oscillators. In this way, the system enables quantum-enhanced sensing in a platform typically considered classical. Our results contribute to a more nuanced understanding of the quantum-to-classical transition and point to new opportunities for exploiting quantum resources in minimally engineered or weakly measured systems.
\section{Two coupled harmonic oscillators}

Interacting systems can be coupled to the environment through entangled eigenstates~\cite{Ciuti2006inputoutputUSC, blais2011dissipation,Bamba2012,
Bamba2013, DeLiberato2014}. In such cases, it becomes impossible to directly measure, detect or destroy the entanglement. To illustrate this, consider two coupled harmonic oscillators described by the Hamiltonian
\begin{align}\label{H}
\hat H =\omega \hat a^\dagger \hat a + \Omega \hat b^\dagger \hat b +\frac{g}{2}\left(\hat a + \hat a^\dagger\right) \left(\hat b + \hat b^\dagger\right),
\end{align}
where $\hat a$ ($\hat b$) annihilates an excitation with energy $\omega$ ($\Omega$), and $g$ is the coupling strength between the oscillators. 
The ground state of this system can be written as~\cite{HongYi1992,emary2003dickemod,Zhou2020}
\begin{align}
\begin{split}\label{eq:squeezedGS}
|\mathrm{GS}\rangle &=\exp\left[\theta\frac{\left(\hat a -\hat a^\dagger\right)\left(\hat b + \hat b^\dagger\right)\omega+\left(\hat a + \hat a^\dagger\right)\left(\hat b- \hat b^\dagger\right)\Omega}{2\sqrt{\omega \Omega}}\right] \times\\
&\quad\times \exp\left[\frac{\xi_+}{2}\left(\hat b^2-\hat b^{\dagger2}\right)\right]\exp\left[\frac{\xi_-}{2}\left(\hat a^2-\hat a^{\dagger2}\right)\right]|0\rangle_a |0\rangle_b,
\end{split}
\end{align}
where the subscripts $a$ and $b$ refer to the eigen bases of the uncoupled harmonic oscillators with frequencies $\omega$ and $\Omega$, respectively. We also introduce a mixing angle (note that we keep $\Omega>\omega$)
\begin{align}
\theta = \frac{1}{2} \tan ^{-1}\left(\frac{2 g \sqrt{\omega \Omega } }{\Omega ^2-\omega ^2}\right),
\end{align}
and two single-mode squeezing parameters
\begin{align}
\xi_- &= -\frac{1}{2}\log \left(\frac{\sqrt{\omega^2 +\Omega^2 -\sqrt{(\omega^2-\Omega^2)^2+4 g^2 \omega \Omega }}}{\sqrt{2} \omega }\right), \\
\xi_+ &= -\frac{1}{2}\log \left(\frac{\sqrt{\omega^2 +\Omega^2 +\sqrt{(\omega^2-\Omega^2)^2+4 g^2 \omega \Omega }}}{\sqrt{2} \Omega }\right).
\end{align}
Note that for $g=\sqrt{\omega \Omega}\equiv g_c $, the squeezing parameter $\xi_-$ becomes infinite which marks the instability threshold.

The above ground state~\eqref{eq:squeezedGS} is clearly entangled. However, when transformed into the energy eigenbasis, the Hamiltonian takes the form of two decoupled harmonic oscillators:
\begin{align}\label{eq:normalmodesH}
\hat H = \omega e^{-2\xi_-} \hat c^\dagger \hat c + \Omega e^{-2\xi_+} \hat d^\dagger \hat d,
\end{align}
with the ground state simply given by
\begin{align}
|\mathrm{GS}\rangle = |0\rangle_c |0\rangle_d,
\end{align}
where subscripts $c$ and $d$ refer to the eigen bases of now coupled harmonic oscillators with {normal} frequencies $\omega e^{-2\xi_-}$ and $\Omega e^{-2\xi_+}$. 
In other words, while the state may appear classical in a particular basis, the system itself can still be fundamentally entangled. At first glance, this may seem paradoxical: can entanglement really be created or removed by a mere change of basis? The answer is more subtle. To speak meaningfully about entanglement, one must first specify a partition of the system into (at least) two subsystems, and this partition is not unique. A change of basis may correspond to a change in this partitioning, transforming what appears as entanglement in one description into a separable state in another, without altering the physical state itself. This perspective is not only a matter of mathematical representation. In relativistic quantum information it is well known that entanglement is {observer dependent}. Different observers may decompose the quantum field into different modes, leading to inequivalent partitions of the Hilbert space and hence to different entanglement structures. For instance, uniformly accelerated observers perceive the Minkowski vacuum as a thermal state (the Unruh effect), which degrades the entanglement seen by inertial observers~\cite{Alsing2003,Fuentes2005,Peres2004}. More generally, relative motion or spacetime curvature can generate or diminish entanglement depending on the chosen partitioning~\cite{Friis2012,MartinMartinez2009}.

Measurement in an arbitrary basis is generally nontrivial. Most quantum systems are not measured directly but rather through coupling to an external environment—often via the electromagnetic field~\cite{Carmichael1993,Wiseman2009}. For systems with a linear energy structure (such as linearly interacting harmonic oscillators), if driven by a coherent state of light, they will emit a coherent state in return, regardless of the complexity of their internal quantum states~\cite{Glauber1963,Gardiner1985,Gardiner2004}. This means that the interacting harmonic oscillators can be coupled to the environment through hybrid entangled eigenmodes described by $\hat c$ and $\hat d$ rather than $\hat a$ and $\hat b$. In the language of the master equation, $\hat c$ and $\hat d$ are jump operators. This suggests that no information about the entanglement of eigenstates is imprinted onto the environment. 

However, this is not entirely true. The quantumness of the system is indirectly encoded in the resonant frequencies of the interacting oscillators. Measuring shifts in these frequencies can, therefore, serve as an entanglement witness. To be specific, the squeezing parameters characterizing the ground state at $T=0$ can be obtained by measuring the resonance frequency since
\begin{align}
    \xi_- &= -\frac12\log{\frac{ \omega_-}{\omega}},\\
    \xi_+ &= -\frac12\log{\frac {\omega_+}{\Omega}},
\end{align}
where $ \omega_- = \omega e^{-2\xi_-}$ and $\omega_+ =  \Omega e^{-2\xi_+}$ are the resonant frequencies of the interacting harmonic oscillators. In other words, although no correlations between mode $a$ and $b$ can be detected directly, the squeezing of the groundstate can be indirectly measured through the new frequencies of modes represented by $ c$ and $ d$, provided that there are no thermal fluctuations.

\section{The inseparability criterion}
Let us have a look now at entanglement. In continuous variable quantum systems, such as optical modes described by canonical quadrature operators, entanglement can be detected via the {inseparability criterion} proposed by Duan \textit{et al.}~\cite{Duan2000} and independently by Simon~\cite{Simon2000}. 
Consider quadrature operators of two bosonic modes, for mode $a$
\begin{align}&\hat{x}_{a} = \frac{1}{{2}}\left(\hat a + \hat a^\dagger\right) \quad \text{and} \quad
\hat{p}_{a} = \frac{1}{{2} i}\left(\hat a - \hat a^\dagger\right),
\end{align}
and analogous for mode $b$. Define the EPR-like operators 
\begin{equation}
    \hat{u} = \hat{x}_a + \hat{x}_b, \qquad \hat{v} = \hat{p}_a - \hat{p}_b.
\end{equation}
Then, for all separable states, the following inequality holds
\begin{equation}
    \Delta^2\hat{u} + \Delta^2\hat{v} \geq 1.
\end{equation}
Therefore, violation of this bound,
\begin{equation}
    \Delta^2\hat{u} + \Delta^2\hat{v} < 1,
\end{equation}
is a sufficient condition for entanglement between modes $a$ and $b$. This criterion is especially useful for Gaussian states and is experimentally accessible via homodyne detection provided that one can measure in the basis of mode $a$ and $b$. 

\begin{figure}[htb!]
    \centering
    \includegraphics[width=0.6\linewidth]{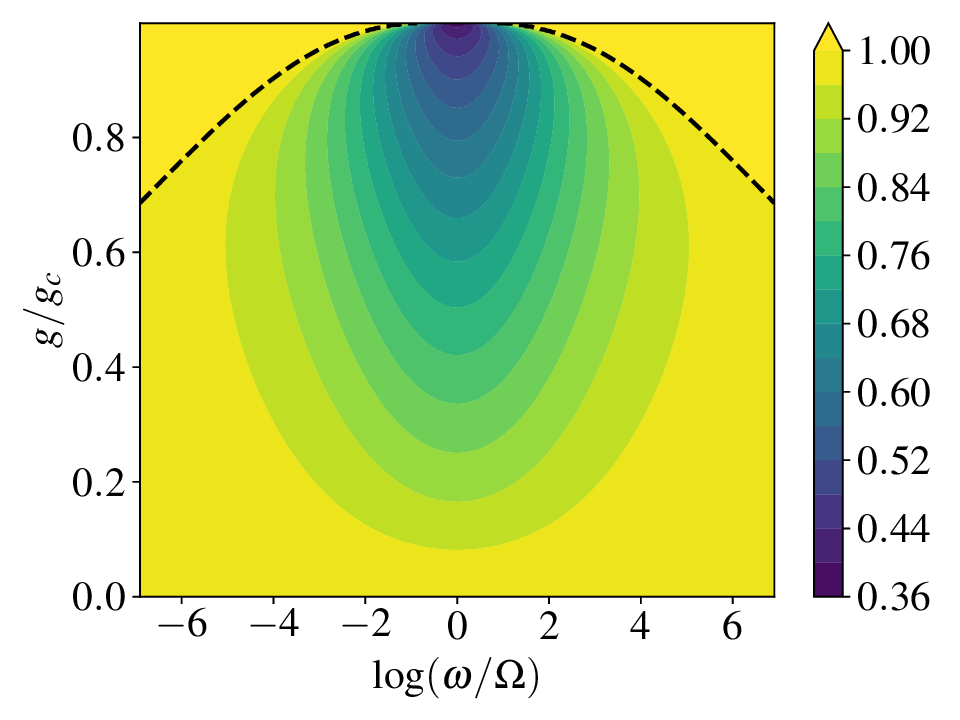}
    \caption{Inseparability criterion at $T=0$ ($\beta=\infty$) versus $\omega/\Omega$ and $g/g_c$ (explicit form given in the Appendix). The scale is truncated at $1$ to highlight the entangled regime (enclosed with a dashed line), with maximal entanglement at $\omega=\Omega$.
}
    \label{fig:fig1}
\end{figure}

This is not the case for interacting systems as the environment is not coupled through the modes $a$ and $b$ but through modes $c$ and $d$. Nevertheless, it is possible to rewrite $\hat u$ and $\hat v$ operators using the modes $c$ and $d$ which for a resonant condition (used for clarity) and calculated for the ground state simplifies to 
\begin{align}
    \frac{e^{-2\xi_-}}{2} + \frac{e^{2\xi_+}}{2} = \frac{\omega_-}{2\omega} + \frac{\omega}{2\omega_+}<1,
\end{align}
which again shows that measuring the frequency shifts can be used to determine the inseparability and entanglement of two interacting harmonic modes. On the other hand, we have the limit $\Omega \gg \omega$ where there should be no entanglement because of the vastly differing energy scales. In this regime one can easily show that the condition for entanglement simplifies to 
\begin{align}
     \frac{e^{-2\xi_-}}{2} + \frac{e^{2\xi_-}}{2} =  \frac{\omega_-}{2 \omega}+\frac{\omega}{2 \omega_- } < 1
\end{align}
which cannot be satisfied. The general condition (see the Appendix) is presented in Fig.~\ref{fig:fig1}. As one could expect, the maximal amount of entanglement arises close to the instability point $g=g_c$ for the resonant case $\omega = \Omega$ (see the Appendix for similar results using logarithimic negativity). 
\section{Finite temperature}
All the discussion above assumes perfect knowledge of the system's temperature. In particular, it assumes that the system is in its ground state. Under this assumption, the frequency measurement can be interpreted as a {conditional witness of entanglement}, valid only at zero temperature. However, in realistic scenarios, the system is typically in a mixed state due to finite thermal excitations, and the question naturally arises: how does thermal noise affect the underlying entanglement and its relation to frequency shifts? To address this, we now consider the case of finite temperature in the resonant case $\omega=\Omega$. The system is then described by a thermal (Gibbs) state as 
\begin{align}
    \hat \rho = \sum_i p_i |e_i \rangle\langle e_i|,
\end{align}
where \(|e_i\rangle\) are the eigenstates of the Hamiltonian~\eqref{eq:normalmodesH} 
and the Boltzmann weights are given by
\begin{align}
    p_i = \frac{e^{-E_i \beta}}{Z}, \quad \text{with}\quad Z = \sum_i e^{-E_i \beta},
\end{align}
where \(E_i\) is the eigenvalue corresponding to \(|e_i\rangle\), \(\beta = 1/T\) is the inverse temperature  (in units where \(k_B = 1\)), and \(Z\) is the partition function ensuring normalization.

\begin{figure}[htb!]
    \centering
   \includegraphics[width=0.6\linewidth]{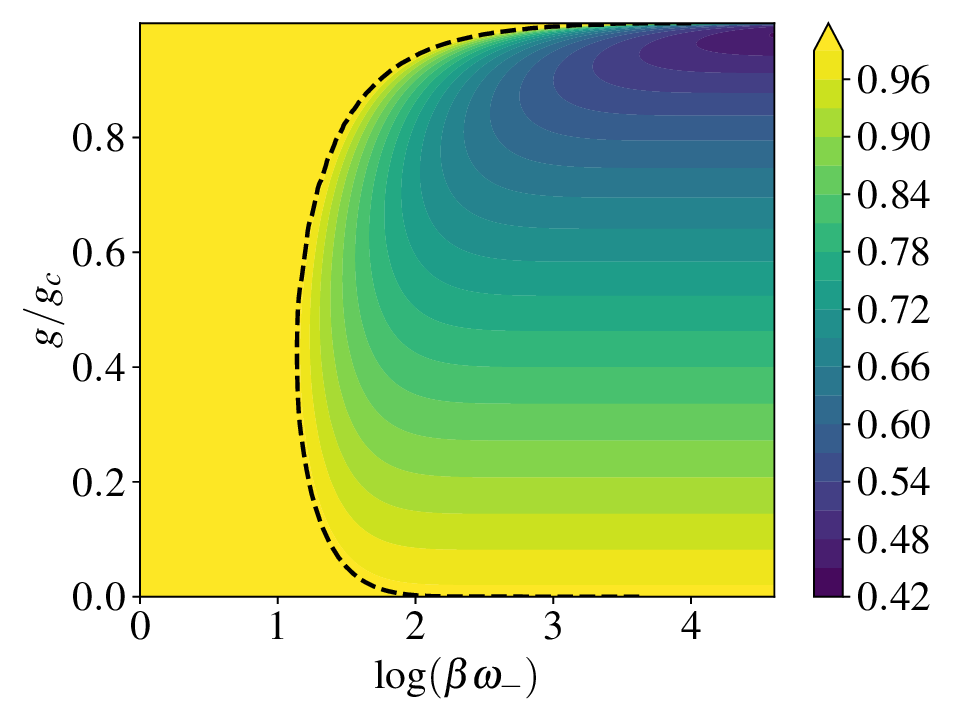}
    \caption{Inseparability criterion as a function of $g/g_c$ and the dimensionless temperature $\beta \omega_-\equiv \omega_-/T$ for the resonant case $\omega=\Omega$. Increasing temperature reduces the entangled region, with entanglement vanishing beyond a specific temperature dependent on $g/g_c$. The scale is truncated at $1$ to highlight the entangled regime (enclosed with a dashed line). Note that $\omega_-$ is a function of $g/g_c$. 
    }
    \label{fig:fig2}
\end{figure}

In order to do it we rewrite the inseparability criterion in the normal mode basis (see the Appendix). After some algebra, it is straightforward to show that for a thermal state, we obtain
\begin{align}
      e^{-2\xi_-}\left(\langle\hat c^\dagger \hat c\rangle+\frac12 \right) + e^{2\xi_+}\left(\langle\hat d^\dagger \hat d \rangle+ \frac12\right)<1,
\end{align}
where $\langle\hat c^\dagger \hat c\rangle$ and $\langle\hat d^\dagger \hat d\rangle$ is the number of thermal excitations in mode $c$ and $d$, respectively:
\begin{align}\label{eq:insT}
    e^{-2\xi_-}\left(\frac{1}{e^{\beta \omega_-}-1}+\frac12 \right) + e^{2\xi_+}\left(\frac{1}{e^{\beta \omega_+}-1}+ \frac12\right)<1,
\end{align}
which confirms that knowing temperature and the frequencies of non-interacting and interacting harmonic oscillators is sufficient to certify entanglement between modes $a$ and $b$. The inseparability criterion from Eq.~\eqref{eq:insT} as a function of the interaction strength and temperature is presented in Fig.~\ref{fig:fig2}. 

The analysis above demonstrates that frequency shifts not only signal entanglement through the inseparability criterion, but also remain quantitatively linked to the degree of two-mode squeezing even at finite temperature. Note that the same microscopic correlations that induce inseparability underlie the macroscopic spectral response of the system. This duality suggests a deeper operational meaning of the observed frequency renormalization. It is not merely a static spectral feature, but a manifestation of the very quantum resources that enable enhanced metrological performance. Consequently, the frequency shifts that witness the entanglement will also carry the metrological advantage. In the following we demonstrate how these hidden correlations—encoded in the system’s structure rather than in its noise statistics—translate into measurable improvements in parameter estimation, revealing a direct bridge between correlations, sensitivity, and frequency shifts.

\section{Quantum enhanced sensitivity through squeezed states}
Let us first revisit the conventional route to quantum-enhanced sensitivity based on squeezing of quantum noise. For this purpose we consider a single quantum harmonic oscillator governed by the Hamiltonian
\begin{align}
\hat H = \omega \hat a^\dagger \hat a.
\end{align}
In frequency metrology, one typically prepares a coherent state by displacing the vacuum~\cite{Glauber1963,Walls1994,Scully1997} and monitors the system’s dynamics to estimate the oscillator frequency~\cite{Giovannetti2004,Paris2009,Pezze2018}. A common choice of measurement is the position quadrature  implemented via homodyne detection~\cite{Leonhardt1997,Gardiner2004}.
\begin{align}
\hat x = \frac{1}{2}(\hat a + \hat a^\dagger),
\end{align}
whose expectation value evolves in time as
\begin{align}
\langle \hat x \rangle = |\alpha| \cos\left(\omega t \right),
\end{align}
where $|\alpha|^2$ is the mean excitation number and $t$ is the evolution time. For a coherent state, the quantum uncertainty in $\hat x$ remains constant:
\begin{align}
\Delta^2 \hat x = \langle \hat x^2 \rangle - \langle \hat x \rangle^2 = \frac{1}{4}.
\end{align}
The quality of the measurement can be quantified by the signal-to-noise ratio,
\begin{align}
S(\omega) = \frac{\mathcal{M}\left( \partial_\omega \langle \hat x \rangle \right)^2}{\Delta^2 \hat x},
\end{align}
where $\mathcal{M}$ (in the following set to 1 for simplicity) is the number of measurements performed to construct the expectation value of $\hat x$.
To improve precision one reduces the quantum noise by using a squeezed coherent state of the form
\begin{align}
\hat D(\alpha) \hat S(\xi) |0\rangle,
\end{align}
where $\hat D(\alpha)$ and $\hat S(\xi)$ are the displacement and squeezing operators, respectively. For such a state, the noise in the $\hat x$ quadrature becomes
\begin{align}
\Delta^2 \hat x = \frac{e^{-2\xi}}{4},
\end{align}
leading to an enhanced signal-to-noise ratio,
\begin{align}
S(\omega) = \frac{\left( \partial_\omega \langle \hat x \rangle \right)^2}{\Delta^2 \hat x} \leq \frac{4|\alpha|^2 t^2}{e^{-2\xi}}.
\end{align}
Here, $\xi$ is the squeezing parameter. For $\xi = 0$, we recover the standard quantum limit, while for $\xi < 0$, the measurement sensitivity is enhanced due to reduced quantum noise.

This is precisely why squeezed states are so valuable in quantum metrology. They allow one to surpass the standard quantum limit by enlarging the signal-to-noise ratio through quantum noise reduction.

\section{Quantum enhanced sensitivity with squeezing the oscillator's frequency}

Now consider a frequency measurement on an interacting harmonic oscillator from Eq.~\eqref{eq:normalmodesH} (here we concentrate on one of the modes only)
\begin{align}
\hat H = \omega e^{{-}2\xi_-}\hat c^\dagger \hat c ,
\end{align}
still with the aim to precisely estimate the bare frequency of this oscillator without interaction $\omega$~\cite{Gietkacxvs-5pb1}. For a coherent state $|\alpha\rangle_c$---squeezed in terms of $\hat a$ and $\hat a^\dagger$ operators---where the subscript $c$ indicates the new basis, it is straightforward to calculate the signal-to-noise ratio. The noise is of course
\begin{align}
    \langle \hat x_c^2\rangle -\langle \hat x_c \rangle^2 = \frac{1}{4},
\end{align}
as it should be for a classical state, but an interesting thing happens with the signal
\begin{align}
    \partial_\omega \langle \hat x_c \rangle = |\alpha|{t}\cos{ \omega_- t} \times \partial_\omega \omega_-.
\end{align}
We find that the signal is multiplied by a derivative of the effective frequency with respect to the frequency of the non-interacting harmonic oscillator that we want to measure. This derivative can be explicitly evaluated. For a sufficiently far off-resonant case ($\omega\ll\Omega$ taken for simplicity) it turns out to be 
\begin{align}
\begin{split}
    \partial_\omega  \omega_- =& \partial_\omega \left[\omega\exp(-2\xi_-)\right]=\partial_\omega  \left[\omega \sqrt{1-g^2/\Omega\omega}\right] =\\=& \exp(2 \xi_-)\left[1-\frac{g^2}{2 \omega  \Omega }\right]>1
    \end{split}
\end{align}
which for large squeezing ($g\sim \sqrt{\omega \Omega}$) simplifies to 
\begin{align}
    \partial_\omega \omega_- = \frac{\exp(2 \xi_-)}2.
\end{align}
As a result, the signal-to-noise ratio becomes
\begin{align}
    S(\omega) = \frac{\left(\partial_\omega \langle \hat x_c\rangle\right)^2}{\Delta^2\hat x_c}\leq \frac{|\alpha|^2 t^2}{\exp(-4\xi_-)}.
\end{align}
Although the system appears entirely classical and no direct squeezing of quantum noise is involved, the signal-to-noise ratio is nonetheless enhanced. The improvement is governed by the squeezing parameter associated with the inaccessible modes $a$ and $b$. In other words, it is possible to obtain a "quantum" enhanced measurement within a seemingly classical system. This highlights that the classical observables of coupled harmonic oscillators are fundamentally shaped by their underlying quantum correlations. Even in the absence of overt quantum features, quantum resources might remain operative and exploitable. Additionally, one could also create squeezing in the new mode $\hat c$ and combine the squeezing effects at two different levels~\cite{PhysRevLett.132.060801,mihailescu2025criticalquantumsensingtutorial}.

\section{Conclusions \& discussion}
We have demonstrated that linearly coupled quantum harmonic oscillators—systems, often regarded as essentially classical due to their Gaussian states and analytical solvability, can harbor hidden quantum correlations. These manifest themselves through measurable, seemingly classical observables. In particular, we have shown that coupling-induced frequency shifts encode two-mode squeezing between the original oscillator modes. From two complementary viewpoints, the frequency shifts and the squeezing describe the same physical effect of coupling, merely expressed in different mode bases. In the normal-mode picture it appears as a frequency renormalization, while in the original basis it manifests as two-mode squeezing. 

Although the underlying squeezing might not be directly observable, it leaves a clear imprint on the system's response, which can be harnessed for quantum-enhanced frequency estimation. Strikingly, this enhancement arises without the need to manipulate quantum noise directly. Instead of relying on traditional noise squeezing to reduce measurement uncertainty, the quantum advantage stems from the implicit squeezing embedded in the eigenstates of the interacting system. This mechanism leads to a signal-to-noise improvement that scales as $\exp(4\xi)$—an enhancement over the standard $\exp(2\xi)$ scaling achieved through explicit squeezing~\cite{Gietkacxvs-5pb1}. Hence in suitable cases, quantum resources might be more efficiently exploited when encoded into the system's structure rather than externally imposed.

Our findings challenge the common held belief that squeezing has no classical counterpart. We propose that the metrological advantage typically attributed to quantum squeezing can in certain systems appear as classical frequency shifts. This suggests a subtle reinterpretation of squeezing. While the underlying correlations are quantum in origin, their effects may be accessed through classical measurements of the system response \cite{Walls1983, Caves1981}. Our results blur the conventional boundary between classical and quantum regimes, showing that systems traditionally considered classical can allow for equal measurement advantages as a quantum model when appropriately interpreted. 

From a wider perspective this also sheds some new light on the classical-to-quantum transition, suggesting that the emergence of quantum-enhanced behavior need not coincide with abrupt changes in nonclassicality indicators; it may instead unfold gradually encoded in how a system responds to parameter variations \cite{Kofler2007, Leggett2002}. In this view, classical observables, such as frequencies, can serve as sensitive witnesses of latent quantum correlations, even in regimes where traditional signatures of quantum-ness are absent \cite{Brunner2014, Adesso2016}. These opens a promising direction for quantum metrology in platforms where direct quantum state control is limited. Weakly interacting or passive systems, which require no state preparation beyond coherent displacements, may still provide access to quantum resources—hidden in plain sight as classical observables \cite{Pezze2018}. Ultimately, this motivates a deeper investigation into how quantum correlations can be embedded, witnessed and utilized across a broad range of physical systems, reframing our understanding of the quantum–classical divide in both conceptual and practical terms.

Let us finally remark, that the results can be naturally extended to interacting spin systems that admit a mapping to a harmonic oscillator, such as the Lipkin–Meshkov–Glick model~\cite{Lipkin1965,LMG2005,LMG2007,LMG2008}. This offers a direct route to certify entanglement—in the form of spin squeezing—by simply monitoring how the energy gap is modified by spin–spin interactions. A more intriguing challenge, however, is to uncover how the energy gap relates to spin squeezing in less symmetric systems that cannot be reduced to a harmonic oscillator description. We anticipate that this connection will hinge on the interplay between the gap and the symmetry properties of the eigenstates~\cite{Gietka2025prep}, potentially opening novel pathways for generating scalable spin squeezing in many-body spin systems. Establishing such a relation would not only broaden the range of experimentally accessible platforms, but could also provide a unifying framework linking spectral properties and metrological performance in complex quantum matter.

\acknowledgements
K. G. would like to thank Adam Miranowicz, Piotr Grochowski, and Johannes Fankhauser for fruitful discussions.





\appendix
\section{Transforming the operators}
The Hamiltonian can be diagonalized by the unitary transformation
\begin{align}
    \hat U 
    &= \exp\!\left[\theta\,\frac{(\hat a - \hat a^\dagger)(\hat b + \hat b^\dagger)\,\omega 
      + (\hat a + \hat a^\dagger)(\hat b - \hat b^\dagger)\,\Omega}{2\sqrt{\omega \Omega}}\right]
      \exp\!\left[\tfrac{\xi_+}{2}\!\left(\hat b^2 - \hat b^{\dagger2}\right)\right]
      \exp\!\left[\tfrac{\xi_-}{2}\!\left(\hat a^2 - \hat a^{\dagger2}\right)\right],
\end{align}
with parameters
\begin{align}
    \theta &= \tfrac{1}{2}\tan^{-1}\!\left(\frac{2 g \sqrt{\omega \Omega}}{\Omega^2 - \omega^2}\right), \\
    \xi_- &= -\tfrac{1}{2}\log \!\left[\frac{\sqrt{\omega^2 + \Omega^2 - \sqrt{(\omega^2-\Omega^2)^2+4 g^2 \omega \Omega}}}{\sqrt{2}\,\omega}\right], \\
    \xi_+ &= -\tfrac{1}{2}\log \!\left[\frac{\sqrt{\omega^2 + \Omega^2 + \sqrt{(\omega^2-\Omega^2)^2+4 g^2 \omega \Omega}}}{\sqrt{2}\,\Omega}\right].
\end{align}

Applying $\hat U$ to the bare annihilation operators gives the Bogoliubov transformations
\begin{align}
   {\hat{c}=}\hat U \hat a \hat U^\dagger 
    &= \frac{\cos\theta}{2\sqrt{\omega \omega_-}}
       \big[(\omega+\omega_-)\hat a + (\omega-\omega_-)\hat a^\dagger\big]
     + \frac{\sin\theta}{2\sqrt{\omega \omega_+}}
       \big[(\omega+\omega_+)\hat b + (\omega-\omega_+)\hat b^\dagger\big], \\
    {\hat{d}=}\hat U \hat b \hat U^\dagger 
    &= \frac{\cos\theta}{2\sqrt{\Omega \omega_+}}
       \big[(\Omega+\omega_+)\hat b + (\Omega-\omega_+)\hat b^\dagger\big]
     - \frac{\sin\theta}{2\sqrt{\Omega \omega_-}}
       \big[(\Omega+\omega_-)\hat a + (\Omega-\omega_-)\hat a^\dagger\big],
\end{align}
with analogous expressions for the Hermitian conjugates. 

Each bare mode is thus mapped to a superposition of creation and annihilation operators of both oscillators. The transformation combines a rotation with single-mode squeezing, thereby providing the metric that consistently relates observables in the bare and normal-mode representations.

\section{General inseparability criterion for the {ground state}} 
In the main text, we focused on the resonant case for the sake of clarity. Here we present the inseparability criterion for the groundstate in a general case expressed just by the frequencies of non-interacting modes $\omega$ and $\Omega$ as well as the frequencies of the interacting modes $\omega_-$ and $\omega_+$. In order to show it, we first derive the condition using the angle $\theta$ which depends on the interaction strength. After some algebra, the condition for the ground state reads 
\begin{align} 
    \frac{(\omega \Omega+\omega_- \omega_+) \left[(\omega+\Omega) (\omega_-+\omega_+)-(\omega_--\omega_+) \left((\omega-\Omega) \cos (2 \theta)-2 \sqrt{\omega\Omega} \sin (2 \theta)\right)\right]}{8 \omega \omega_- \Omega \omega_+} <1.
\end{align}
By using the definition of the mixing angle $\theta$, we arrive at
\begin{align}
    \frac{(\omega \Omega + \omega_- \omega_+) \left[(\omega+\Omega) (\omega_-+\omega_+)+\frac{(\omega_--\omega_+) \left(-4 \omega \Omega g-(\omega-\Omega)^2 (\omega+\Omega)\right)}{\sqrt{\left(\omega^2-\Omega^2\right)^2+4 \omega \Omega g^2}}\right]}{8 \omega \omega_- \Omega \omega_+}<1.
\end{align}
By expressing the interaction strength through the frequencies
\begin{align}
    g =\sqrt{\frac{\left(\omega_-^2-\omega_+^2\right)^2-\left(\omega^2-\Omega^2\right)^2}{4 \omega \Omega}},
\end{align}
we can rewrite the inseparability criterion just with 4 frequencies:
\begin{align}
\frac{(\omega \Omega+\omega_- \omega_+) \left[(\omega+\Omega) (\omega_-+\omega_+)-
\frac{2 \sqrt{\omega \Omega} \sqrt{{\left(\omega_-^2-\omega_+^2\right)^2-\left(\omega^2-\Omega^2\right)^2}}+(\omega+\Omega) (\omega-\Omega)^2}{\omega_-+\omega_+}\right]}{8 \omega \omega_- \Omega \omega_+}<1.
\end{align}

\section{Logarithmic negativity for two coupled harmonic oscillators}
\label{app:logneg}

Since the ground state of the system is completely determined by its normal-mode and non-interacting frequencies, any measure of quantum correlations, including entanglement, can be expressed directly in terms of these frequencies. To quantify the bipartite entanglement between two bosonic modes, we use now the \emph{logarithmic negativity}~\cite{PhysRevA.65.032314,PhysRevLett.95.090503}, which is one of the most commonly employed entanglement measures for Gaussian states due to its computational simplicity and clear operational meaning.  

For a two-mode Gaussian state described by the covariance matrix $\boldsymbol{\sigma}$, the logarithmic negativity is defined as  
\begin{equation}
E_{\mathcal N} = \max\!\big[0,\; -\log_2(2\tilde{\nu}_-)\big],
\end{equation}
where $\tilde{\nu}_-$ denotes the smallest \emph{symplectic eigenvalue} of the \emph{partially transposed} covariance matrix.  
This quantity captures the degree to which the partial transposition operation---analogous to time reversal of one mode---violates the physicality condition of the covariance matrix.  
The threshold for separability is given by \(2\tilde{\nu}_-=1\):
\begin{itemize}
    \item If \(2\tilde{\nu}_- \ge 1\), the state is separable (no entanglement).
    \item If \(2\tilde{\nu}_- < 1\), the state is entangled, and the smaller the value of \(\tilde{\nu}_-\), the stronger the entanglement.
\end{itemize}

\subsection*{Model and canonical quadratures}

We consider the bilinear Hamiltonian used in the main text,
\begin{equation}
\hat H = \omega\,\hat a^\dagger \hat a + \Omega\,\hat b^\dagger \hat b
+ \frac{g}{2}\big(\hat a + \hat a^\dagger\big)\big(\hat b + \hat b^\dagger\big),
\end{equation}
describing two harmonic oscillators of frequencies $\omega$ and $\Omega$ coupled via a position–position interaction of strength $g$.  Introducing canonical quadratures (setting $\hbar=1$),
\begin{align}
\hat X_a &= \frac{1}{\sqrt{2\omega}}(\hat a + \hat a^\dagger), &
\hat P_a &= i\sqrt{\frac{\omega}{2}}(\hat a^\dagger - \hat a), \\
\hat X_b &= \frac{1}{\sqrt{2\Omega}}(\hat b + \hat b^\dagger), &
\hat P_b &= i\sqrt{\frac{\Omega}{2}}(\hat b^\dagger - \hat b),
\end{align}
we recover the canonical commutation relations $[\hat X_j,\hat P_k]=i\delta_{jk}$.  
In these coordinates, the Hamiltonian becomes
\begin{equation}
\hat H = \frac{1}{2}\big(\hat P_a^2 + \hat P_b^2\big)
+ \frac{1}{2}\big(\omega^2\hat X_a^2 + \Omega^2\hat X_b^2\big)
+ g\sqrt{\omega\Omega}\,\hat X_a\hat X_b.
\end{equation}
It is convenient to write the position part as $\tfrac{1}{2}\mathbf{X}^T V\mathbf{X}$, where $\mathbf{X}=(\hat X_a,\hat X_b)^T$ and
\[
V = \begin{pmatrix}
\omega^2 & g\sqrt{\omega\Omega} \\[4pt]
g\sqrt{\omega\Omega} & \Omega^2
\end{pmatrix},
\]
while the momentum part remains diagonal, $\tfrac{1}{2}\mathbf{P}^T\mathbf{P}$.

\subsection*{Normal modes and diagonalization}

To find the normal modes, we diagonalize the symmetric matrix $V$.  
Its eigenvalues correspond to the squared normal-mode frequencies:
\begin{equation}
\omega_\pm^2 = \frac{\omega^2 + \Omega^2}{2}
\pm \sqrt{\left(\frac{\omega^2 - \Omega^2}{2}\right)^2 + g^2\omega\Omega}.
\end{equation}
The associated orthogonal transformation $R(\theta)$ corresponds to a rotation by an angle $\theta$, determined by
\begin{equation}
\tan(2\theta) = \frac{2g\sqrt{\omega\Omega}}{\Omega^2 - \omega^2}= \sqrt{\frac{\left(\omega_-^2-\omega_+^2\right)^2}{\left(\omega ^2-\Omega ^2\right)^2}-1}.
\end{equation}
Introducing the rotated quadratures
\begin{equation}
\begin{pmatrix}\hat X_+ \\ \hat X_- \end{pmatrix} = R(\theta)^T\begin{pmatrix}\hat X_a \\ \hat X_b \end{pmatrix}, \qquad
\begin{pmatrix}\hat P_+ \\ \hat P_- \end{pmatrix} = R(\theta)^T\begin{pmatrix}\hat P_a \\ \hat P_b \end{pmatrix},
\end{equation}
the Hamiltonian takes the simple form
\begin{equation}
\hat H = \frac{1}{2}\sum_{s=\pm}\big(\hat P_s^2 + \omega_s^2\hat X_s^2\big),
\end{equation}
i.e. two independent harmonic oscillators with frequencies $\omega_\pm$.

\subsection*{Ground-state covariances}

The ground state of each normal mode is a vacuum with covariances
\begin{equation}
\langle \hat X_s^2\rangle = \frac{1}{2\omega_s}, \qquad
\langle \hat P_s^2\rangle = \frac{\omega_s}{2}, \qquad s=\pm.
\end{equation}
There are no cross-correlations between position and momentum in this basis.  
Transforming back to the local quadratures yields the covariance blocks (with $c=\cos\theta,\ s=\sin\theta$):
\begin{align}
X_{11} = c^2\frac{1}{2\omega_+} + s^2\frac{1}{2\omega_-},\quad 
X_{22} = s^2\frac{1}{2\omega_+} + c^2\frac{1}{2\omega_-},\quad
X_{12} = cs\left(\frac{1}{2\omega_+} - \frac{1}{2\omega_-}\right), \label{eq:Xentries}
\end{align}
\begin{align}
P_{11} = c^2\frac{\omega_+}{2} + s^2\frac{\omega_-}{2}, \quad
P_{22} = s^2\frac{\omega_+}{2} + c^2\frac{\omega_-}{2}, \quad
P_{12} = cs\left(\frac{\omega_+}{2} - \frac{\omega_-}{2}\right).\label{eq:Pentries}
\end{align}
Thus, the full covariance matrix in the ordering $(\hat X_a,\hat X_b,\hat P_a,\hat P_b)$ is block-diagonal,  
$\boldsymbol{\sigma} = \mathrm{diag}(\boldsymbol{X},\boldsymbol{P})$.

\subsection*{Partial transpose and reduction to a $2\times2$ problem}

Partial transposition with respect to mode $b$ acts in phase space as $\hat P_b\mapsto -\hat P_b$.  
This operation leaves the position block unchanged but flips the sign of $P_{12}$.  
Denoting the modified momentum block by $\widetilde{\boldsymbol{P}}$, the partially transposed covariance matrix is $\widetilde{\boldsymbol{\sigma}}=\boldsymbol{X}\oplus\widetilde{\boldsymbol{P}}$. For block-diagonal covariances of this type, the squares of the symplectic eigenvalues of $\widetilde{\boldsymbol{\sigma}}$ coincide with the eigenvalues of the $2\times2$ matrix
\begin{equation}
M = \boldsymbol{X}\,\widetilde{\boldsymbol{P}}.
\end{equation}
Let $\mu_\pm$ be the eigenvalues of $M$ ($\mu_+\ge\mu_-$).  
Then the partially transposed symplectic eigenvalues are $\tilde{\nu}_\pm = \sqrt{\mu_\pm}$, and the relevant one for entanglement is the smaller, $\tilde{\nu}_-=\sqrt{\mu_-}$.

\subsection*{Closed form for $\mathrm{tr}\,M$ and $\mu_-$}

Using Eqs.~\eqref{eq:Xentries}--\eqref{eq:Pentries}, and recalling that the element $P_{12}$ changes sign under partial transposition, one obtains
\begin{equation}
\mathrm{tr}\,M = X_{11}P_{11} + X_{22}P_{22} - 2 X_{12}P_{12}.
\end{equation}
It is worth noting that the specific choice of dimensionless or rescaled quadratures (as defined in the main text) does not affect this expression, nor the subsequent results.  
After some straightforward algebra, the above relation can be simplified to the compact form
\begin{equation}
\mathrm{tr}\,M = \frac{1}{4\omega_+\omega_-}
\Big[\omega_+^2 + \omega_-^2 - \cos^2(2\theta)\,(\omega_+ - \omega_-)^2\Big].
\end{equation}
For the pure ground state, the determinant factorizes as 
\(\det M = (\det\boldsymbol{X})(\det\widetilde{\boldsymbol{P}})\),
which evaluates to \((1/2)^4 = 1/16\).  
Consequently, the smaller eigenvalue of \(M\) is given by
\begin{equation}
\mu_- = \frac{1}{2}
\left[\mathrm{tr}\,M - 
\sqrt{(\mathrm{tr}\,M)^2 - \tfrac{1}{16}}\right],
\end{equation}
from which the smallest symplectic eigenvalue follows as
\begin{equation}
\tilde{\nu}_- = \sqrt{\mu_-}.
\end{equation}

\subsection*{Resonant case: simple analytical expressions}

In the resonant situation, $\omega=\Omega$, the diagonalization simplifies considerably.  
The mixing angle becomes $\theta=\pi/4$, for which $\cos(2\theta)=0$.  
The normal-mode frequencies reduce to
\begin{equation}
\omega_\pm^2 = \omega^2 \pm g\omega, \qquad\text{i.e.}\qquad
\omega_\pm = \omega\sqrt{1 \pm \frac{g}{\omega}}.
\end{equation}
Substituting these relations into the above expressions gives
\begin{equation}
\mathrm{tr}\,M = \frac{1}{4}\left(\frac{\omega_+}{\omega_-} + \frac{\omega_-}{\omega_+}\right),
\end{equation}
and with $\det M = 1/16$, one obtains the particularly transparent result
\begin{equation}
\tilde{\nu}_- = \frac{1}{2}\sqrt{\frac{\omega_-}{\omega_+}}.
\end{equation}
The logarithmic negativity then follows as
\begin{equation}
E_{\mathcal N} = \frac{1}{2}\log_2\!\left(\frac{\omega_+}{\omega_-}\right)
= \frac{1}{2}\log_2\!\left(\sqrt{\frac{1+g/\omega}{1-g/\omega}}\right).
\end{equation}
These expressions show that the entanglement grows with increasing coupling $g$, and diverges as $g\to\omega$, where the lower normal-mode frequency softens to zero, signaling the approach to an instability.

\subsection*{Finite temperature}

If the two normal modes are thermally populated with mean occupations $n_\pm = (e^{\beta\omega_\pm}-1)^{-1}$, the variances $\langle \hat X_s^2\rangle$ and $\langle \hat P_s^2\rangle$ acquire the multiplicative factors $(2n_\pm+1)$.  
Repeating the rotation and partial transpose steps yields, in the resonant case,
\begin{equation}
\tilde{\nu}_-(T) = \frac{1}{2}\sqrt{\frac{2n_-+1}{2n_++1}\cdot\frac{\omega_-}{\omega_+}},
\end{equation}
and the thermal logarithmic negativity is then
\begin{equation}
E_{\mathcal N}(T) = \max\!\big(0,\; -\log_2(2\tilde{\nu}_-(T))\big).
\end{equation}
As expected, thermal excitations degrade the entanglement, which vanishes when $2\tilde{\nu}_-(T)\ge1$.


\end{document}